\documentstyle [12pt,a4p,epsfig,amsmath,multicol]{article}
\newcommand{\CA}{{\rm C}$_{A}$ }
\newcommand{\CB}{{\rm C}$_{B}$ }
\newcommand{\CAP}{{\rm C}$_{A'}$ }
\newcommand{\CBP}{{\rm C}$_{B'}$ }

\begin{document}
\vspace*{0.6cm}

\begin{center} 
{\normalsize\bf Clock rates, clock settings and the physics of the space-time
  Lorentz transformation}
\end{center}
\vspace*{0.6cm}
\centerline{\footnotesize J.H.Field}
\baselineskip=13pt
\centerline{\footnotesize\it D\'{e}partement de Physique Nucl\'{e}aire et 
 Corpusculaire, Universit\'{e} de Gen\`{e}ve}
\baselineskip=12pt
\centerline{\footnotesize\it 24, quai Ernest-Ansermet CH-1211Gen\`{e}ve 4. }
\centerline{\footnotesize E-mail: john.field@cern.ch}
\baselineskip=13pt
\vspace*{0.9cm}
\abstract{ A careful study is made of the operational meaning of the time symbols 
  appearing in the space-time Lorentz transformation. Four distinct symbols,
  with different physical meanings, are needed to describe reciprocal
   measurements involving stationary and uniformly-moving clocks. Physical
  predictions concern only the observed rate of a clock as a function of its relative speed,
 not its setting. How the failure to make this distinction leads to the conventional
  predictions of spurious `relativity of simultaneity' and `length contraction' 
 effects in special relativity is explained.}
\vspace*{0.9cm}
\normalsize\baselineskip=15pt
\setcounter{footnote}{0}
\renewcommand{\thefootnote}{\alph{footnote}}

    The purpose of this paper is to discuss the physical consequences, for measurements of 
    space and time intervals, of the Lorentz transformation\footnote{The Lorentz Transformation (1)-(4)
   was first given by Larmor in 1900~\cite{Larmor}. The same transformation
  appears in 1904 in Lorentz's last pre-relativity paper~\cite{Lorentz}. The latter paper was cited
  in 1905 by Poincar\'{e}~\cite{Poincare1905} who first introduced the appellation `Lorentz Transformation'.
  Einstein
 independently derived the transformation in the same year~\cite{Ein1}. See  Ref.~\cite{Kittel} for
  a discussion of historical priority issues}(LT):
\begin{eqnarray}
    x' & = & \gamma[x-vt] \\
    t' & = & \gamma[t-\frac{v x}{c^2}] \\
    y'  & = & y \\
   z'  & = & z
\end{eqnarray}
 where $\gamma \equiv 1/\sqrt{1-(v/c)^2}$ and $c$ is the speed of light in vacuum.
  For this it is necessary to establish the exact operational correspondence between the mathematical
   symbols representing space-time events ($x$,$y$,$z$,$t$) and  ($x'$,$y'$,$z'$,$t'$) in two inertial
   frames S and S' and actual clock readings and measured spatial intervals. In the LT (1)-(4) the Cartesian
    spatial axes of S and S' are parallel and
  the frame S' is in uniform motion, relative to S along the common $x$, $x'$ axis,  with velocity $v$.
  \par In order apply the LT to any experiment, the times $t$ and $t'$ must be identified with those
   recorded by clocks at rest in either S or S'. The simplest experiment that can be conceived
   will use just one clock in each frame, say, C and C'. Using these clocks two different, but 
   reciprocal, experiments may be performed. The two clocks may be viewed from S, or they may be viewed
    from S'. In order to describe these experiments, four physically distinct time symbols are required:
     $\tau$, $t'$, $\tau'$ and  $t$. The symbols $\tau$ ($\tau'$) represent the times recorded by the stationary 
    clocks C (C') as observed in S (S'), while the symbols $t$ ($t'$) represent the times recorded by the moving 
    clocks C (C') as observed from  S'(S).
   \par To analyse these experiments it will be found convenient to use the invariant interval 
     relations~\cite{Poincare1906,Mink} connecting  $\tau$,$t'$ or  $\tau'$,$t$ that may be derived from the LT (1)-(4),
 or its inverse: 
  \begin{equation}
      c^2  (\Delta t')^2 -  (\Delta x')^2
      = c^2  (\Delta \tau)^2 -  (\Delta x)^2 
   \end{equation}    
  \begin{equation}
      c^2  (\Delta t)^2 -  (\Delta x)^2
      = c^2  (\Delta \tau')^2 -  (\Delta x')^2 
   \end{equation} 
   Eqn(5) describes the experiment where C' is viewed from S, so that so that the transformed events 
   lie on the world line of C'. Since C' is at rest in S', $\Delta x' = 0$, while its equation
   of motion in S is: $\Delta x = v \Delta \tau$. Thus (5) may be written as

  \[   c^2  (\Delta t')^2  = c^2 (1-\frac{v^2}{c^2})(\Delta \tau)^2 \]
  
    or
   \begin{equation}
     \Delta \tau = \gamma \Delta t' 
    \end{equation}
    Similarly, (6) describes the experiment where C is viewed from S', and the transformed
   events lie on the world line of C. In this case  $\Delta x = 0$, and  $\Delta x' = - v \Delta \tau'$
   so that Eqn(6) gives:
   \begin{equation}
     \Delta \tau'= \gamma \Delta t  
    \end{equation}
  The equations (7) and (8) describe the relativistic time dilatation (TD) effect. In may be succinctly
    stated as follows:
   \begin{equation}
  \frac{{\rm rate~of~moving~clock}}{{\rm rate~of~stationary~clock}} =\frac{\Delta t'}{\Delta \tau}
     = \frac{\Delta t}{\Delta \tau'} =\frac{1}{\gamma}
   \end{equation}
   This relation  holds for clocks of any construction whatever, and is, as first emphasised by
   Einstein, of a purely kinematical nature.
   \par In order to discuss length interval measurements in two different inertial frames, at
     least four clocks are needed, say, \CA and \CB, at rest in S, and \CAP and \CBP at rest
    in S'. Suppose that  \CA and \CB lie along the $x$-axis, separated by the distance $L$. The
    separation of  \CAP and \CBP,  $L'$, is chosen in such a way that when the $x$-coordinates
    of \CA and \CAP coincide, so do those of \CB and  \CBP. At this instant, suppose that the readings
     of the clocks in the frame S are: $\tau_0$(\CA), $\tau_0$(\CB), $t'_0$(\CAP)  and  $t'_0$(\CBP), with
    corresponding spatial coordinates:
    \[x_0({\rm C}_A) = x_0({\rm C}_{A'}) = 0,~~x_0({\rm C}_B) = x_0({\rm C}_{B'}) = L,~~x'_0({\rm C}_{A'})
   =0,~~ x'_0({\rm C}_{A'}) = L' \] 
     In an abbreviated notation the LT for events on the world lines of \CAP and \CBP are:
  \begin{eqnarray}
   x'_{A'} & = & \gamma[ x_{A'} -v \Delta \tau_A] = 0 \\  
\Delta t'_{A'} & = & \gamma[\Delta \tau_A  - \frac{v  x_{A'}}{c^2}] \\
    x'_{B'}- L' & = & \gamma[ x_{B'} -L -v \Delta \tau_B] = 0 \\  
\Delta t'_{B'} & = & \gamma[\Delta \tau_B  - \frac{v(x_{B'} -L)}{c^2}]
 \end{eqnarray}
  where, for example $x_{A'} \equiv x$(\CAP) and $\Delta \tau_A \equiv \tau$(\CA)$-\tau_0$(\CA). 
   Note that in (12) $L$ is the value of $x_{B'}$ when $\Delta \tau_B = 0$, and is independent of the value of $v$.  
   Eliminating $x_{A'}$ between (10) and (11) and  $x_{B'}$ between (12) and (13) enables (10)-(13) to be written
   equivalently as:
  \begin{eqnarray}
    x_{A'} & = & v \Delta \tau_A = \gamma \beta \Delta t'_{A'} \\
    \Delta \tau_A & = & \gamma \Delta t'_{A'} \\
 x_{B'}-L & = & v \Delta \tau_B = \gamma \beta \Delta t'_{B'} \\
    \Delta \tau_B & = & \gamma \Delta t'_{B'}
 \end{eqnarray}
  Since events with clock settings  $t'_0$(\CAP) and $t'_0$(\CBP) are simultaneous, and the clocks run at the same rate,
  then $t'$(\CAP) and $t'$(\CBP) are also simultaneous provided that:
  \begin{equation}
   t'(C_{A'})- t'(C_{B'}) = t'_0(C_{A'})-t'_0(C_{B'})
  \end{equation}
  or, rearranging:
    \begin{equation}
      \Delta t'_{A'} = \Delta t'_{B'} \equiv \Delta t'
   \end{equation}
  Considering now simultaneous events in S', and in virtue of the identity: $\gamma^2 -\gamma^2 \beta^2 = 1$, 
  (14),(15) and (16),(17) are pairs of parametric equations for hyperbolae in the $\Delta \tau$ versus $x$ plane:
   \begin{equation}
 c^2(\Delta \tau_A)^2- ( x_{A'})^2 = c^2(\Delta t')^2 =  c^2(\Delta \tau_B)^2- ( x_{B'}-L)^2
 \end{equation}      
    Since (15), (17) and (19) require that $\Delta \tau_A = \gamma \Delta t' = \Delta \tau_B$, it follows from (20) that,
    at any instant in S:
     \begin{equation}
x_{B'}(\beta) - x_{A'}(\beta) = L
 \end{equation} 
  where the $\beta$ dependence of $x_{A'}$ and $x_{B'}$ following from(14) and (16) at fixed $\Delta t'$ is explicitly
  shown. Since $x \rightarrow x'$  as $\beta \rightarrow 0$, it follows from (21) that
     \begin{equation}
x_{B'}(0) - x_{A'}(0) = x'_{B'} - x'_{A'} \equiv L' =  L
 \end{equation}
   The spatial separation of the clocks is therefore a Lorentz invariant quantity
   that has the same value in all inertial frames --there is no `relativistic length contraction' (LC).
   How this spurious effect arises in the standard interpretation, in special relativity, of the 
   LT will be explained below.
     \par In the calculations above, no particular clock synchronisation procedure was introduced.
     Similar results will now be rederived using the LT
   with a particular choice of spatial coordinates and `clock synchronisation convention'.
   The latter concept will be first explained. A clock or a watch is intended to record
   time intervals. Its precision depends on how well the rate of the clock stays constant
   as a function of elapsed time. This requires understanding and control of the physical
   processes underlying the operation of the clock. Clocks and watches usually have a knob 
   by which the actual time indicated may be `set'. This setting process may be performed
   in an arbitary manner by the clock owner, but is completely unrelated to the
   physics of the clock mechanism which determines only the {\it rate} at which the clock
   runs. Sometimes, as in the case of a stopwatch, there may
   also be a button that simply resets the indicated time to zero. As discussed above,
   when the LT is used to predict the results of space-time experiments, at least one
   moving and one stationary clock is required. In general, the initial times displayed
   by these clocks can be set to arbitary values without affecting any physical predictions.
   For concreteness, suppose that both clocks are stopped (i.e. no longer record time) and the moving
   and stationary ones are set to the values $t'_0$ and $\tau_0$ respectively. As previously,
   the moving clock is confined to the $x$-axis of the frame S. When the $x$-coordinates
   of the clocks coincide they are both started (i.e. begin to record time). The two
   numbers $t'_0$ and $\tau_0$ define a particular `clock synchronisation convention'(CSC).
    All predictions for physical effects are independent of this convention. This method to synchronise
    two clocks in different inertial frames was introduced by Einstein in the original
    special relativity paper~\cite{Ein1} and was termed `external synchronisation' by Mansouri 
    and Sexl~\cite{MS}.
    \par It is easily seen that the LT of Eqns (1) and (2) above, cannot be used to implement
    a general CSC as described above. Setting $x = x' = 0$, when the clocks in S and S' have the same
    $x$-coordinate, then (1) gives $t \equiv \tau = 0$,  so that, in consequence, (2) gives $t' = 0$.
    This means
    that the particular CSC:  $t'_0 = \tau_0 = 0$ is built into the LT (1)-(2) at $x = x' = 0$. It is like
    a stop watch that can be reset, but not set! Worse, as will be shown below, it can only
    be reset at one particular spatial location. The modification of the LT (1)-(2) to enable
    the implementation of a general CSC (allowing it to describe, not only a stopwatch,
    but also a watch with knobs that enable to set the clock hands at any chosen positions) is straightforward.
    Eqns(1) and (2) are replaced by:
\begin{eqnarray}
    x' & = & \gamma[x-v(\tau-\tau_0)] \\
    t'-t'_0 & = & \gamma[\tau-\tau_0-\frac{v x}{c^2}] 
\end{eqnarray}
    where, as discussed above, the symbol $\tau$ is used for the time indicated by the clock a rest in S, 
  and $t'$ for the
  time indicated by the clock at rest in S', as viewed for S.
    \par The importance of introducing the time offsets $t_0'$ and $\tau_0$ to correctly describe
     synchronised clocks at different spatial positions (equivalent to placing additive constants
     on the right sides of Eqns(1) and (2)) was already pointed out by Einstein in the 
     original 1905 special relativity paper\footnote{See P40, P134 respectively, in the English translations in Ref.~\cite{Ein1}.}:
 \par {\tt If no assumption whatever be made as to the initial position of the \newline moving system
   and as the the zero point of $\tau$,} ($t'$ in the notation used above) \newline 
  {\tt  an additive constant
   is to be placed on the right side of each of these} \newline~(the LT (1)-(4)){\tt equations.}
     \par To the present writer's best knowledge, this important remark was never taken into
     account by Einstein, or anybody else, before the work reported in~\cite{JHFLLT}. 
    \par At the expense of adding the two parameters
  $\tau_0$ and $t'_0$ the general CSC is now applied to the clocks when their $x$-coordinates
   coincide at $x = x' = 0$. However, if the clocks are somewhere else (say at $x = x' = L$), or if
   it is wished to use a different coordinate system to specify the position of the clocks, but to use
    the same CSC, (23) and (24) will not work, any more than does the LT (1) and (2). The solution, 
    as in going from (1)-(2) to (23)-(24), is to make the replacements $\tau \rightarrow \tau -\tau(L)$
    and  $t' \rightarrow t' -t'(L)$ in (23) and (24) where the constants $\tau(L)$ and $t'(L)$ are now to
    be chosen in such a way that the two clocks at $x = x' = L$ will have the same CSC as the clocks at
     $x = x' = 0$ described by (23) and (24). For this $\tau(L)$ and $t'(L)$ must satisfy the equations:
 \begin{eqnarray} 
    L & =  & \gamma [L+ v\tau(L)]  \\
   t'(L) & = & \gamma [\tau(L) +\frac{v L}{c^2}] 
\end{eqnarray}
    The solution of (25) and (26) is
 \begin{eqnarray} 
   \tau(L) & = & -\frac{L}{\gamma v}(\gamma -1) \\
   t'(L) & = & \frac{L}{\gamma v}(\gamma -1) = - \tau(L)
\end{eqnarray}
 With the substitutions $\tau \rightarrow \tau -\tau(L)$, $t' \rightarrow t' -t'(L)$
 (23) and (24) give
\begin{eqnarray}
    x' & = & \gamma[x-v(\tau-\tau_0-\tau(L))] \\
    t'-t'_0- t'(L) & = & \gamma[\tau-\tau_0-\tau(L)-\frac{v x}{c^2}] 
\end{eqnarray}
 or, in virtue of (27) and (28),
\begin{eqnarray}
    x'- L & = & \gamma[x-L-v(\tau-\tau_0)] = 0 \\
    t'-t'_0 & = & \gamma[\tau-\tau_0-\frac{v(x-L)}{c^2}] 
\end{eqnarray}
  This is the LT appropriate for a clock pair at $x = x' = L$ that
  has the same CSC as the clock pair at  $x = x' = 0$ that is described by (23) and (24).
  \par Introducing the relative coordinates:
 \begin{eqnarray}
    \hat{x} & \equiv &  x - x_{C'}(\tau = \tau_0) \\
    \hat{x}' & \equiv &  x' - x'_{C'} 
\end{eqnarray}
  where  $x_{C'}$ and $ x'_{C'}$ are the coordinates of C' in S and S' respectively,
   both (23),(24) and (31),(32) may be written as:
\begin{eqnarray}
     \hat{x}'  & = & \gamma[ \hat{x}-v(\tau-\tau_0)] = 0 \\
    t'-t'_0 & = & \gamma[\tau-\tau_0-\frac{v \hat{x}}{c^2}] 
\end{eqnarray}
 which is identical to (23) and (24), or to the original LT (1) and (2), when $t'_0 = \tau_0 = 0$. except
  that the origin of the coordinate system in S' always coincides with the position
   of the clock C', whatever its spatial position. The LT (35)-(36) has been called
  elsewhere a `local' LT~\cite{JHFLLT}.
 \par Using (36) to eliminate $\hat{x}$ from (35) yields:
   \begin{equation}
  \tau - \tau_0 = \gamma (t' - t'_0)
\end{equation}
  which is equivalent to Eqn(7) above. (35) and (36) then predict the same TD effect for any pair of clocks 
    with the same CSC, independently of their spatial positions. 
 \par  The equations of motion of the clocks \CAP and \CBP in the frame S in (35) are the same as in 
  Galilean relativity. Since the correlated LC and relativity of simultaneity
  effects do not exist (the calculational error from which they originate is explained 
  below), the sole modification of the physics of flat space-time resulting from the LT in the present example
  is to replace the relation
    \begin{equation}
\tau - \tau_0 =  t' - t'_0
   \end{equation}
 expressing the absolute (Newtonian) character of time in Galilean relativity, by the universal, translationally invariant,
 TD relation (37). 

  \par The origin of the spurious `relativity of simultaneity'' and `length contraction' effects
   of conventional special relativity theory will now be explained. The LT (1)-(2) corresponds
    to the CSC $\tau_0 = t'_0 = 0$ for a clock \CAP situated at the $x'$-origin in S'. 
     If exactly the same LT is applied to a clock \CBP at $x' = L'$ at $t \equiv \tau = 0$ the
    following equations are obtained:
\begin{eqnarray}
 L' & = & x'({\rm C}_{B'}) = \gamma  x({\rm C}_{B'}) \\
 t'_0 & \equiv & t'({\rm C}_{B'}) = -\frac{\gamma vx({\rm C}_{B'})}{c^2} = -\frac{ v L'}{c^2}   
\end{eqnarray}
 Thus the CSC for \CBP is $\tau_0 = 0$, $ t'_0 = - v L'/c^2$
 instead of  $\tau_0 = t'_0 = 0$ when the same LT is applied at $x' = 0$. As mentioned
 earlier, the resetable-to-zero stopwatch described by the LT (1)-(2) at $x' = 0$ no longer
  works when  $x' \ne 0$. In standard special relativity theory, to date, this trivial difference
  in the synchronisation convention given by the LT (1)-(2) when  $x' \ne 0$ has been interpreted
   as a real physical effect --`relativity of simultaneity' (RS). Introducing the clock \CAP 
   at $x' = 0$ for which $ x({\rm C}_{A'}) =  x'({\rm C}_{A'}) = 0$ at $\tau = 0$, Eqn(39)
   may be written as:
    \begin{equation}
  L' =  x'({\rm C}_{B'})- x'({\rm C}_{A'}) = \gamma [ x({\rm C}_{B'})- x({\rm C}_{A'})] \equiv  \gamma L
   \end{equation}
   This is the spurious LC effect  --the distance between the clocks is
   predicted to be shorter by the fraction $1/\gamma$ in the frame S in which they are in motion.
   It is also clear by considering the motion of similarly accelerated objects in a single frame
   of reference~\cite{JHFLLT} that the LC effect of
   Eqn(41) is not a genuine physical effect. Both it and the RS effect
   of Eqn(40) are the result of misinterpretation of the time symbols in Eqns(1) and (2).
   These represent clock settings which are, intially, fixed at $t = t' =0$ by the condition $x = x' = 0$,
   but which may, in general, be freely chosen to have any values when $x = x' = 0$,
   as in Eqns(23) and (24). They are
   not times recorded by synchronised clocks {\it at any values of $x$ and $x'$}, as assumed in the conventional
   interpretation of Eqns(40) and (41). The only {\it physical} prediction of the LT concerning moving
   clocks is the TD effect of Eqn(9). The spurious RS and LC effects are the result of a confusion
   between arbitary clock settings and physical time intervals. Physics determines only the {\it rates} of
   clocks. Their {\it settings} are arbitary and have no physical significance.
   \par Even though it is clear from 
   the above that the RS effect predicted by conventional special relativity theory
   results from an erroneous argument, its existence (or non-existence) is still
   experimentally testable. Some specific satellite-borne experiments have recently been proposed to
    search for the RS effect~\cite{JHFLLT,JHFRSE}. At the time of writing, there is ample experimental
     verification of TD but none of RS or LC~\cite{JHFLLT}. 
 
  \par{\bf Acknowledgements}
  
   \par Discussions with, or correspondence from: G.Boas, B.Echenard,
  M.Gr\"{u}newald, Y.Keilman, M.Kloster, B.Rothenstein and D.Utterback, are gratefully acknowledged.
  Although they may not (yet)    
  agree with conclusions of this paper, their questions and critical remarks were very useful to focus
  attention on the essential issues to be addressed.

\end{document}